\documentclass[twoside]{dis07}
\usepackage[latin1]{inputenc}
\usepackage[dvips]{graphicx,epsfig,color}
\usepackage{wrapfig,rotating}
\usepackage{amssymb,amsmath,array}

\begin{document}
\title{Beyond Collins and Sivers: further measurements of the target
transverse spin-dependent azimuthal asymmetries in semi-inclusive DIS
from COMPASS}

\author{Aram Kotzinian
%
\thanks{On leave from Yerevan Physics Institute, 375036 Yerevan,
Armenia and JINR, 141980 Dubna, Russia.}~
  \\on behalf of the COMPASS collaboration
%
\vspace{.3cm}\\
%
INFN, Sezione di Torino, Via P. Giuria 1, I-10125 Torino, Italy\\
 \\
}

\maketitle

\begin{abstract}
In semi-inclusive DIS of polarized leptons on a transversely
polarized target eight azimuthal modulations appear in the
cross-section. Within QCD parton model four azimuthal asymmetries
can be interpreted at leading order, two of them being the already
measured Collins and Sivers asymmetries. The other two leading twist
asymmetries, related to different transverse momentum dependent
quark distribution functions, and also additional four asymmetries
which can be interpreted as twist-three contributions have been
measured for the first time at COMPASS, using a 160 GeV/c
longitudinally polarized ($P_{beam}\simeq -0.8$) muon beam and a
transversely polarized $^6LiD$ target. We present here the
preliminary results from the 2002-2004 data.

\end{abstract}

\section{Introduction}
During last few years many exciting experimental results and theory
development are obtained in SIDIS on the transversely polarized
target. Up to now only the measurements~\cite{herm,comp1,comp2} of
Sivers and Collins asymmetries were performed by HERMES and COMPASS
collaborations and together with data from BELLE~\cite{belle} they
allow a first extraction the transversity and Sivers transverse
momentum dependent (TMD) distribution functions (DFs) and Collins
fragmentation function (FF). In addition to these, the general
expression of SIDIS cross section~\cite{Kotzinian:1994dv} contains
six more target transverse polarization dependent azimuthal
asymmetries. Here we present the preliminary results on these
asymmetries for the first time measured by COMPASS from the
2002-2004 data.

\subsection{Definition of asymmetries \label{sec:asym_def}}

In the following the notations of Ref.~\cite{Bacchetta:2006tn} are
used. There are eight azimuthal modulations related to the target
transverse polarization:
\begin{eqnarray}
&&w_1(\phi_h, \phi_s)=\sin(\phi _h -\phi _s ),\;
w_2(\phi_h,\phi_s)=\sin(\phi _h +\phi _s ),\; w_3(\phi_h,
\phi_s)=\sin(3\phi _h -\phi _s ),\nonumber \\
&& w_4(\phi_h, \phi_s)=\sin(\phi _s ),\; w_5(\phi_h,
\phi_s)=\sin(2\phi _h -\phi _s ),\; w_6(\phi_h, \phi_s)=\cos(\phi _h
-\phi _s ),\\
 && w_7(\phi_h, \phi_s)=\cos(\phi _s ),\; w_8(\phi_h,
\phi_s)=\cos(2\phi _h -\phi _s ),\nonumber
\end{eqnarray}
where first two correspond to Sivers and Collins effects. The
expression for the cross section in interest can be represented as
\begin{eqnarray}\label{eq:cros-sect-short}
    d \sigma(\phi_h, \phi_s, ...) & \propto & (1 +
    |{\bf S}_T| {\sum_{i=1}^5} D^{w_i(\phi_h, \phi_s)} A_{UT}^{w_i(\phi_h,
\phi_s)}
    w_i(\phi_h,\phi_s)\\ \nonumber
    & + &
    P_{beam} |{\bf S}_T| {\sum_{i=6}^8} D^{w_i(\phi_h, \phi_s)}
    A_{LT}^{w_i(\phi_h, \phi_s)} w_i(\phi_h,\phi_s) + ...\big),
\end{eqnarray}
where ${\bf S}_T$ is the target transverse polarization. We factored
out the explicitly calculable depolarization factors,
$D^{w_i(\phi_h, \phi_s)}$, and defined the asymmetries as the ratios
of corresponding structure functions to unpolarized one:

\begin{equation}\label{eq:as_def}
    A_{BT}^{w_i(\phi_h, \phi_s)} \equiv  \frac{F_{BT}^{w_i(\phi_h,
    \phi_s)}}{F_{UU,T}},
\end{equation}
where $B=L$ or $B=U$ corresponds to beam polarization dependent or
independent part of asymmetry.

The depolarization factors entering in Eq.
(\ref{eq:cros-sect-short}) depend only on $y$ and are given as
\begin{eqnarray}\label{eq:depol}
    && D^{\sin(\phi _h -\phi _s )}(y) = 1,\;\;
    D^{\cos(\phi _h -\phi _s )}(y) =\frac {y(2-y)} {1+(1-y)^2},\nonumber\\
    && D^{\sin(\phi _h +\phi _s )}(y) = D^{\sin(3\phi _h +\phi _s )}(y)
=\frac {2(1-y)} {1+(1-y)^2}, \nonumber\\
    && D^{\sin(2\phi _h -\phi _s )}(y) = D^{\sin(\phi _s )}(y) = \frac
{2(2-y)\sqrt{1-y}} {1+(1-y)^2}, \\
    &&D^{\cos(2\phi _h -\phi _s )}(y) = D^{\cos(\phi _s )}(y) = \frac
{2y\sqrt{1-y}}
    {1+(1-y)^2}.\nonumber
\end{eqnarray}

The asymmetries extracted from the data as amplitudes of
corresponding azimuthal modulations (raw asymmetries) are then given
by
\begin{eqnarray}\label{eq:as_exp}
A_{UT, \; raw}^{w_i(\phi_h, \phi_s)} &=&D^{w_i(\phi_h, \phi_s)}(y) f
|S_T| A_{UT}^{w_i(\phi_h, \phi_s)}\, , \;\;(i=1,5),\\
A_{LT, \; raw}^{w_i(\phi_h, \phi_s)} &=&D^{w_i(\phi_h, \phi_s)}(y) f
P_{beam} |S_T| A_{LT}^{w_i(\phi_h, \phi_s)}\, , \;\;(i=6,8),
\end{eqnarray}
where $f$ is the target polarization dilution factor.

In the QCD parton model the asymmetries $A_{LT}^{\cos (\phi _h -\phi
_s )}$ and $A_{UT}^{\sin (3\phi _h -\phi _s )}$ are given by the
ratio of convolutions of spin-dependent to spin-independent twist
two DFs and FFs, for example
\begin{equation}\label{eq:examp_h1tp}
    A_{UT}^{\sin (3\phi _h -\phi _s )} =  \frac{h_{1T}^{\perp\,q} \otimes H_{1q}^{\bot h}}
    {f_1^q \otimes D_{1q}^{h}},
\end{equation}
and can be used for extraction of DFs $g_{1T}^q$ and
$h_{1T}^{\perp\,q}$ describing the quark longitudinal and transverse
(along quark transverse momentum) polarization in the transversely
polarized nucleon. The other asymmetries can be interpreted as Cahn
kinematic corrections to spin effects on the transversely polarized
nucleon~\cite{Kotzinian:1994dv}, for example:
\begin{equation}
    A_{LT}^{\cos (\phi _s )} =  \frac{M}{Q} \frac{g_{1T}^q \otimes
D_{1q}^h}{f_1^q \otimes D_{1q}^{h}}.
\end{equation}

\section{Results} \label{sec:results}
\begin{figure}
\centerline{\includegraphics[width=1.\columnwidth]{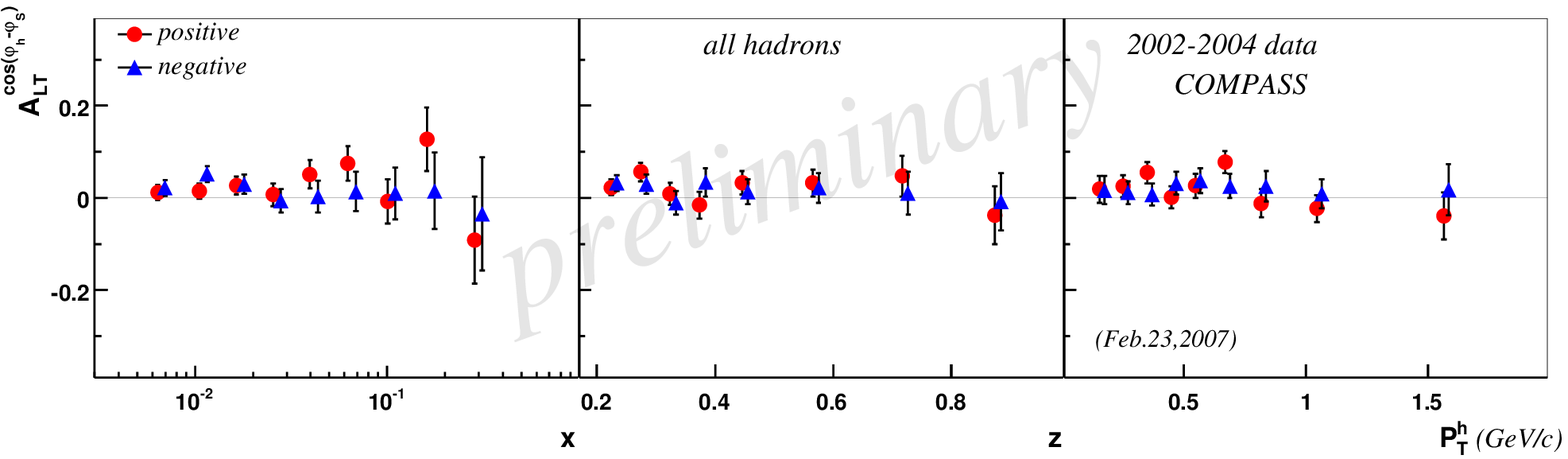}}
\centerline{\includegraphics[width=1.\columnwidth]{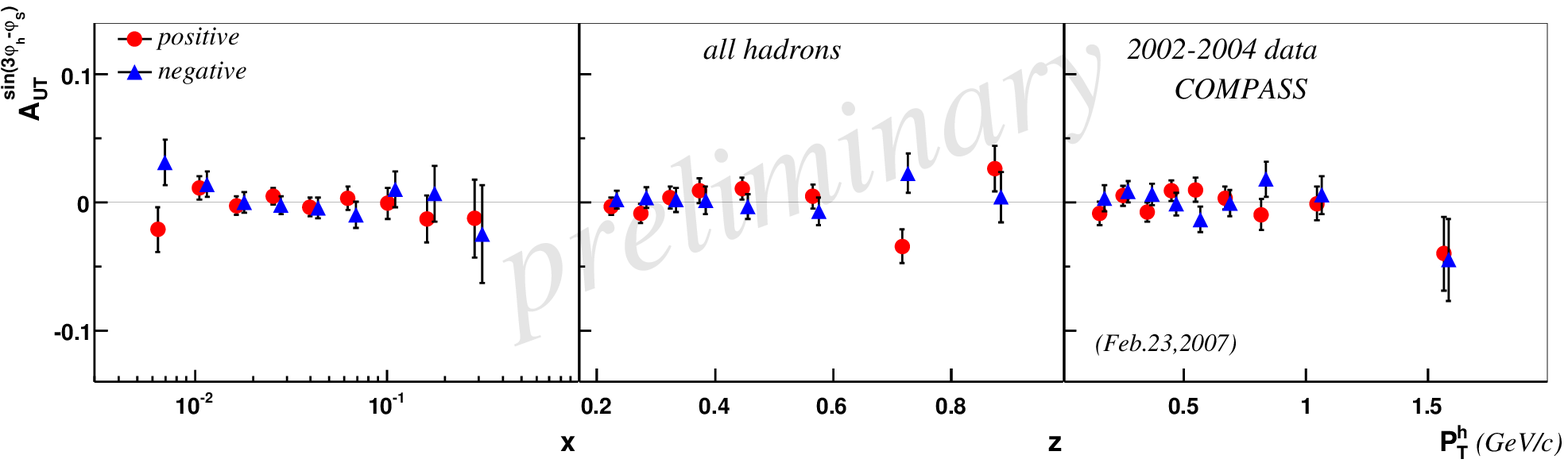}}
\caption{The asymmetries $A_{LT}^{\cos (\phi _h -\phi _s )}$ and
$A_{UT}^{\sin (3\phi _h -\phi _s )}$ as a function of $x$, $z$ and
$P_T^h$.}\label{Fig:1}
\end{figure}

\begin{figure}
\centerline{\includegraphics[width=1.\columnwidth]{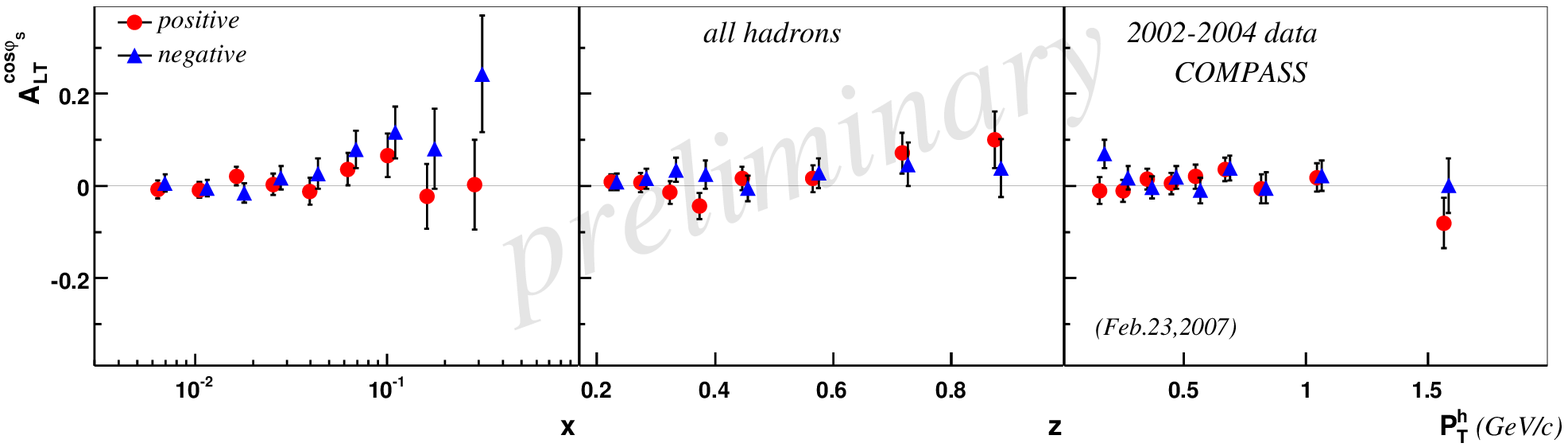}}
\centerline{\includegraphics[width=1.\columnwidth]{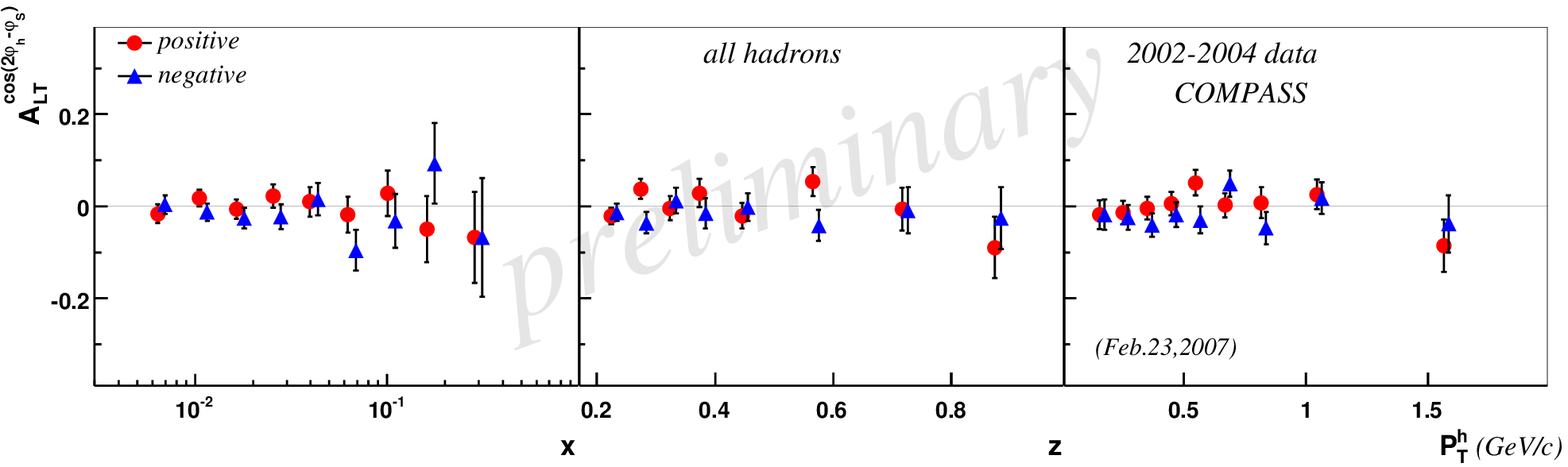}}
\centerline{\includegraphics[width=1.\columnwidth]{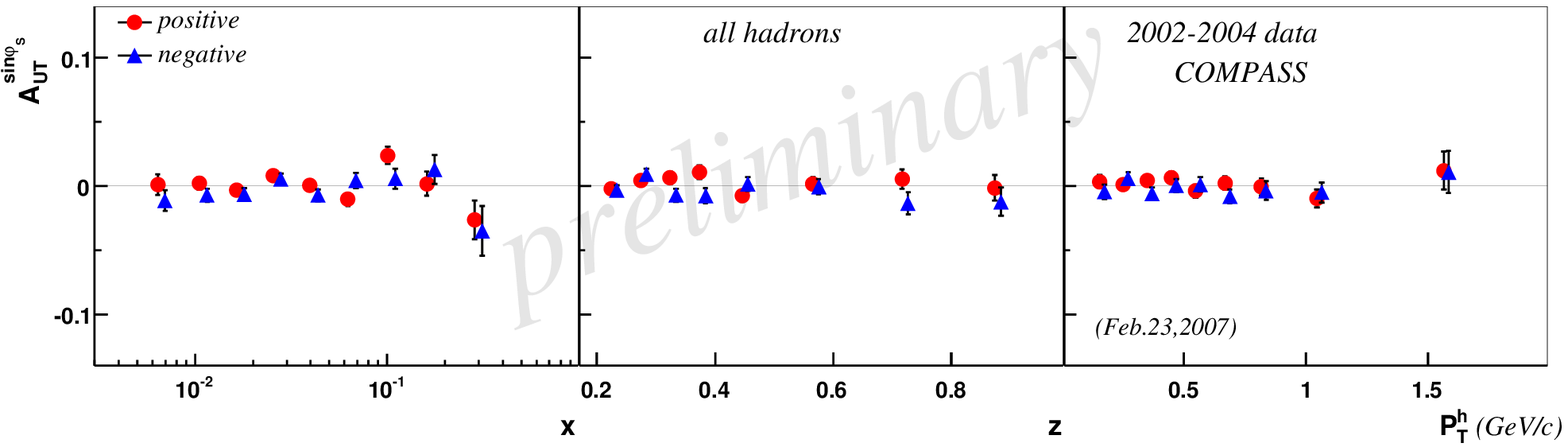}}
\centerline{\includegraphics[width=1.\columnwidth]{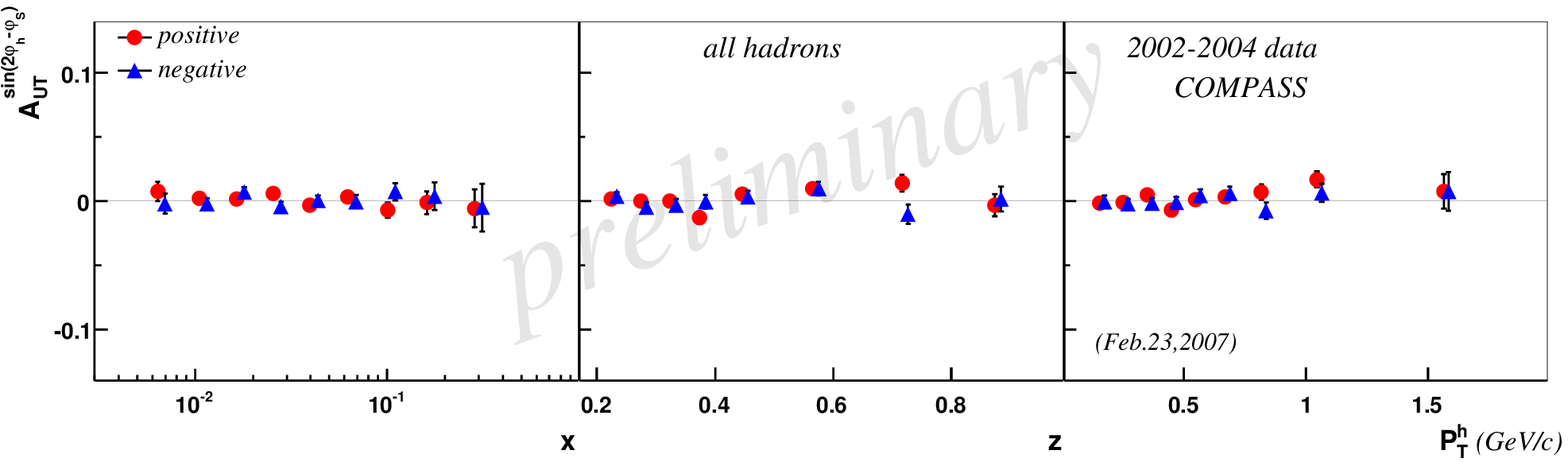}}
\caption{The asymmetries $A_{LT}^{\cos \phi _s}$, $A_{LT}^{\cos
(2\phi _h -\phi _s )}$, $A_{UT}^{\sin \phi _s }$ and $A_{UT}^{\sin
(2\phi _h -\phi _s )}$ as a function of $x$, $z$ and
$P_T^h$.}\label{Fig:2}
\end{figure}

The event selection and asymmetry extraction are done as described
in~\cite{comp2}. The following kinematic cuts were imposed: $Q^2>1$
(GeV/c)$^2$, $W>5$ GeV, $0.1<y<0.9$, $P_T^h>0.1$ GeV/c and $z>0.2$.
In Figs. \ref{Fig:1} and \ref{Fig:2} for the first time we present
six target transverse spin dependent asymmetries extracted from
COMPASS 2002--2004 data collected on deuterium target. The estimated
systematic errors are smaller than statistical. All six newly
measured asymmetries are compatible with zero within statistical
errors.

\begin{footnotesize}

\end{footnotesize}

\end{document}